\documentclass[times, twocolumn]{aastex63}
\usepackage{CJK}
\usepackage{graphicx}
\usepackage{enumerate}
\usepackage{amssymb, amsmath}
\usepackage{natbib}
\usepackage{color}
\usepackage{ulem}
\usepackage{dblfnote}
\usepackage{appendix}
\usepackage{hyperref}
\usepackage[stable]{footmisc}
\usepackage{comment}
\usepackage{footnote}

\usepackage[nodayofweek]{datetime}
\newdateformat{monthyearday}{%
  \THEYEAR\ \monthname[\THEMONTH] \twodigit{\THEDAY}}
  
\submitjournal{AAS Journals}
\shorttitle{Keck/NIRC2 J0337}
\shortauthors{Uyama et al.}

\begin{document}

\title{A spatially-resolved large cavity of the J0337 protoplanetary disk in Perseus}

\correspondingauthor{Taichi Uyama}
\email{tuyama@ipac.caltech.edu}

\author[0000-0002-6879-3030]{Taichi Uyama}
    \affiliation{Infrared Processing and Analysis Center, California Institute of Technology, 1200 E. California Blvd., Pasadena, CA 91125, USA}
    \affiliation{NASA Exoplanet Science Institute, Pasadena, CA 91125, USA}
    \affiliation{National Astronomical Observatory of Japan, 2-21-1 Osawa, Mitaka, Tokyo 181-8588, Japan}

\author[0000-0003-4769-1665]{Garreth Ruane}
    \affiliation{Jet Propulsion Laboratory, California Institute of Technology, 4800 Oak Grove Dr., Pasadena, CA 91109, USA}


\author[0000-0002-6964-8732]{Kellen Lawson}
    \affiliation{Department of Physics and Astronomy, University of Oklahoma, Norman, OK 73019, USA}

\author{Takayuki Muto}
    \affiliation{Division of Liberal Arts, Kogakuin University
2665-1, Nakano-cho, Hachioji-chi, Tokyo, 192-0015, Japan}


\author[0000-0002-5627-5471]{Charles Beichman}
    \affiliation{NASA Exoplanet Science Institute, Pasadena, CA 91125, USA}
    \affiliation{Infrared Processing and Analysis Center, California Institute of Technology, 1200 E. California Blvd., Pasadena, CA 91125, USA}
    
\author[0000-0003-2458-9756]{Nienke van der Marel}
    \affiliation{Leiden Observatory, Niels Bohrweg 2, 2333 CA Leiden, The Netherlands}
    \affiliation{Physics \& Astronomy Department, University of Victoria, 3800 Finnerty Road, Victoria, BC, V8P 5C2, Canada}

\begin{abstract}
We present Keck/NIRC2 $K_{\rm p}L_{\rm p}$ high-contrast imaging observations of a J0337 protoplanetary disk. The data discover the spatially-resolved large cavity, which is the second report among protoplanetary disks in the Perseus star forming region after the LkH$\alpha$~330 system. Our data and forward modeling using RADMC-3D suggests $\sim80$~au for the cavity radius. There is discrepancy between J0337's SED and the modeled SED at $\sim10\micron$ and this suggests an unseen inner disk. We also searched for companions around J0337 but did not detect any companion candidates at separations between $0\farcs1$ and $2\farcs5$. The $L_{\rm p}$-band detection limit corresponds to $\sim20 M_{\rm Jup}$ at 60~au, $\sim9-10 M_{\rm Jup}$ at 90~au, and $\sim3 M_{\rm Jup}$ at $>120$~au. Compared with other young systems with large cavities such as PDS~70 and RX~J1604, multiple Jovian planets, a single eccentric Jovian planet, or a massive brown-dwarf at an inner separation  could exist within the cavity.
\end{abstract}


\section{Introduction} \label{sec: Introduction}

Perseus is one of the nearby star forming regions \citep[$\sim300$ pc derived by Gaia and VLBA observations;][]{Ortiz-Leon2018} and has a young age \citep[$\sim1-3$ Myr for major clusters of IC~348 and NGC~1333;][]{Luhman2016}. 
Spectroscopic studies have reported some possible transitional disks in Perseus - for example, \cite{vanderMarel2016} used Spitzer photometry and IRS data and identified several dozens of transitional disk candidates, some of which potentially have large cavities ($>$50 au).  Transition disk cavities have been linked to both photoevaporative clearing as part of the evolutionary disk dissipation process \citep{Alexander2014} and to the clearing by massive protoplanets which are already forming in the disk \citep{Lin1979}.
Previous observational studies suggested Jovian planets in the transitional disk with cavity $>50$ au \citep[e.g. RX J1604, PDS 70;][]{Dong2017,Hashimoto2012} and eventually VLT/SPHERE high-contrast imaging observations reported the first convincing protoplanet in the PDS~70 disk \citep{Keppler2018}.
However, the number of convincing protoplanets embedded in protoplanetary disks is still small and it is important for better understandings of planet formation mechanisms to detect/characterize more protoplanets. 

2MASS~J03370363+3039291 (hereafter J0337) is a member of the Perseus star forming region, although slightly isolated from IC~348 and NGC~1333, with an IR excess and \cite{vanderMarel2016} suggested a large gap in its disk from the archival Spitzer catalog. 
However, other than the gap information a limited number of stellar/disk parameters are known in this system because this system has not been prioritized compared with other YSOs in the major clusters.
Here we present Keck/NIRC2 high-contrast imaging observations to report that our data affirmed the large cavity as predicted in \cite{vanderMarel2016}.
Section \ref{sec: Observations and Data reduction} describes our observations and data reduction. In Section \ref{sec: Results} the result of the post-processing is presented. Finally we present our SED fitting and forward modeling results and discuss the gap opening mechanisms in the J0337 disk in Section \ref{sec: Discussions}.

\section{Observations and Data Reduction} \label{sec: Observations and Data reduction}
\subsection{Observations} \label{sec: Observations}

J0337 was chosen from the sample of new transition disk candidates with large cavities from \cite{vanderMarel2016}, based on SED analysis of hundreds of Spitzer-selected YSOs. J0337 was selected based on its optical brightness (PI: Nienke van der Marel).
On UT 2016 October 15 we observed J0337 by Keck/NIRC2 $L_{\rm p}$ band (3.776 \micron) combined with the vector vortex mask and with the total exposure time of 2120 seconds, providing the parallactic angle change of $91\fdg33$.
After confirming the disk feature (see Section \ref{sec: Results}) we conducted follow-up observations for this target using NIRC2 $K_{\rm p}$ band (2.124 \micron) on UT 2021 February 2 without a coronagraph mask (PI: Garreth Ruane, backup target). 
We achieved 2100 seconds and 38\fdg94 for the total exposure time and the parallactic angle change.
Both observations were conducted under the vertical angle mode for angular differential imaging \citep[ADI;][]{Marois2006} to suppress the stellar halo and instrumental speckles. The typical full-width at half maximum (FWHM) of the point spread function (PSF) measured 9 pix ($\sim90$~mas) at $L_{\rm p}$ band and 5 pix ($\sim50$~mas) at $K_{\rm p}$ band, respectively.
We note that no reference-PSF stars for reference-star differential imaging \citep[RDI; e.g.][]{Ruane2019} were observed in both epochs and we apply only ADI for post-processing. 

\subsection{Data Reduction} \label{sec: Data Reduction}

First we pre-processed (bad pixel correction, flat fielding, sky subtraction, and image registration) the obtained NIRC2 raw frames \citep[see also][]{Ruane2019}. 
We then performed ADI-based post-processing to subtract the stellar halo by producing the most likely reference PSFs combined with the {\tt pyKLIP} algorithms\footnote{\url{https://pyklip.readthedocs.io/en/latest/index.html}} \citep{pyklip} that utilized Karhunen-Lo\`eve Image Projection \citep[KLIP;][]{Soummer12}.
We adopted minrot=10$^\circ$ in the {\tt pyklip} setting, which is not the most aggressive parameters for the ADI reduction, so that we can avoid a severe self-subtraction effect that distorts the geometry and attenuates the flux of the disk features within $\rho\leq0\farcs3$ detected in our observations (see Section \ref{sec: Results} for the ADI results).

\section{Results} \label{sec: Results}

\begin{figure*}
\begin{tabular}{cc}
\begin{minipage}{0.5\hsize}
    \centering
    \includegraphics[width=0.8\textwidth]{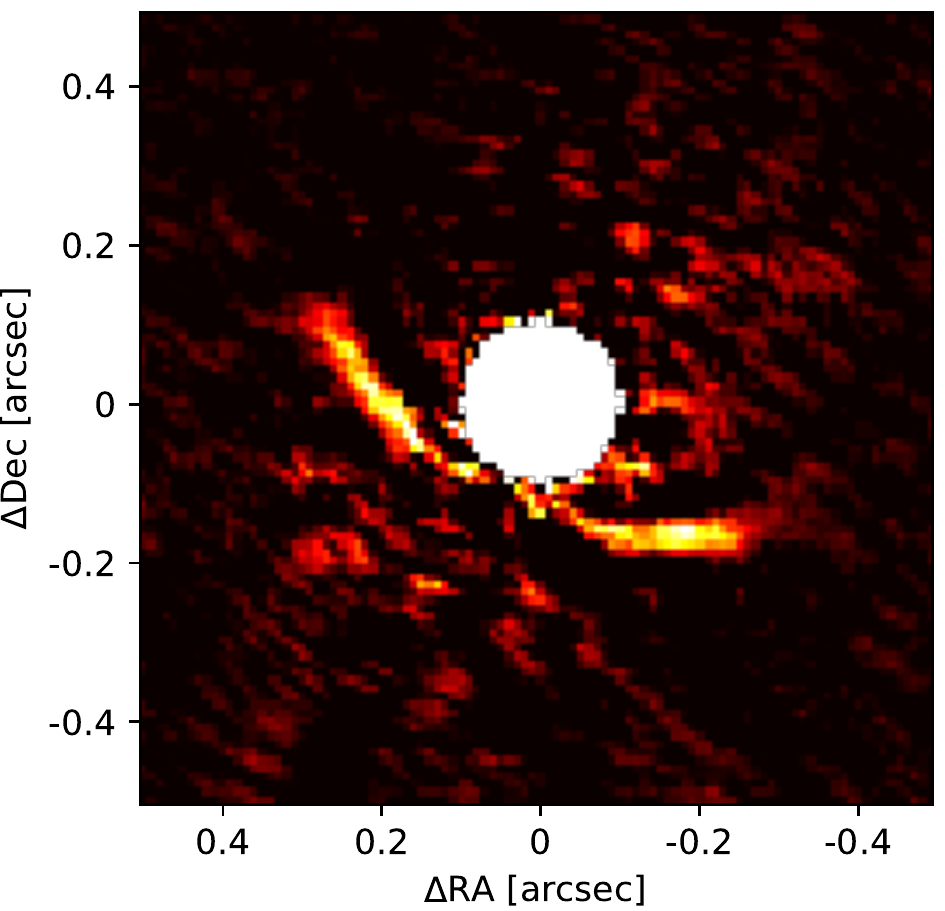}
\end{minipage}
\begin{minipage}{0.5\hsize}
    \centering
    \includegraphics[width=0.8\textwidth]{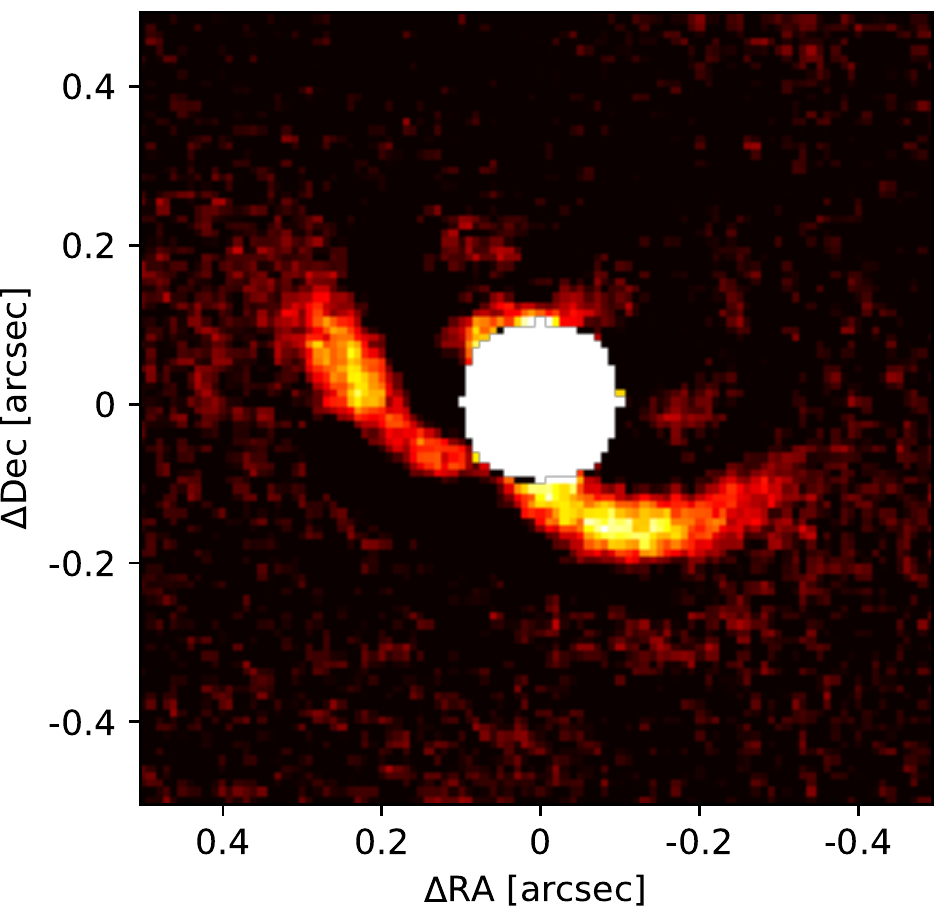}
\end{minipage}
\end{tabular}
\caption{Keck/NIRC2 observations of J0337 at $K_{\rm p}$ band (left) and $L_{\rm p}$ band (right). KL=4 is adopted for both images. The central star is masked by the algorithms. North is up and East is left.}
    \label{fig: NIRC2 result}
\end{figure*}

In the both epochs we confirmed the same arc-like feature (see Figure \ref{fig: NIRC2 result}) with a signal to noise ratio (SNR) $\sim4-5$ for its spine, which indicates that the arc feature is gravitationally bound to J0337.
This arc feature is likely to correspond to forward-scattered light from an inner edge of the outer disk around J0337. The backward-scatter components are not detected in our observations.
Considering the SNR and the aggressiveness for the ADI reduction to avoid severe self-subtraction that attenuates the disk feature, we adopted KL=4 to present our results. However, note that this non-aggressive post-processing still leaves self-subtraction. We conducted injection test by burying artificial point sources at a variety of separations and position angles, which resulted in $>50\%$ self-subtraction particularly in the $K_{\rm p}$-band data occurs within $0\farcs3$ because of its small field rotation angle.
The inner dark region indicates a cavity of $\sim0\farcs25$ in radius, which is consistent with the prediction from the SED fitting study \citep[$r_{\rm in}=50^{+30}_{-10}$~au;][]{vanderMarel2016}.
From the geometry of the arc feature we roughly estimate an inclination and a position angle to be $\sim60^\circ$ and $\sim25^\circ$ respectively. The gap size is estimated by forward modeling (see Section \ref{sec: Forward Modeling}).
Our high-contrast imaging results show the second large cavity in the Perseus transitional disks after the LkH$\alpha$~330 system where a large cavity and a pair of spirals were reported in its disk \citep[][]{Isella2013,Akiyama2016,Uyama2018}. 


\section{Discussions} \label{sec: Discussions}
\subsection{SED Fitting for Stellar Parameters} \label{sec: SED Fitting}

\begin{figure*}
\begin{tabular}{cc}
\begin{minipage}{0.6\hsize}
    \centering
    \includegraphics[width=1\textwidth]{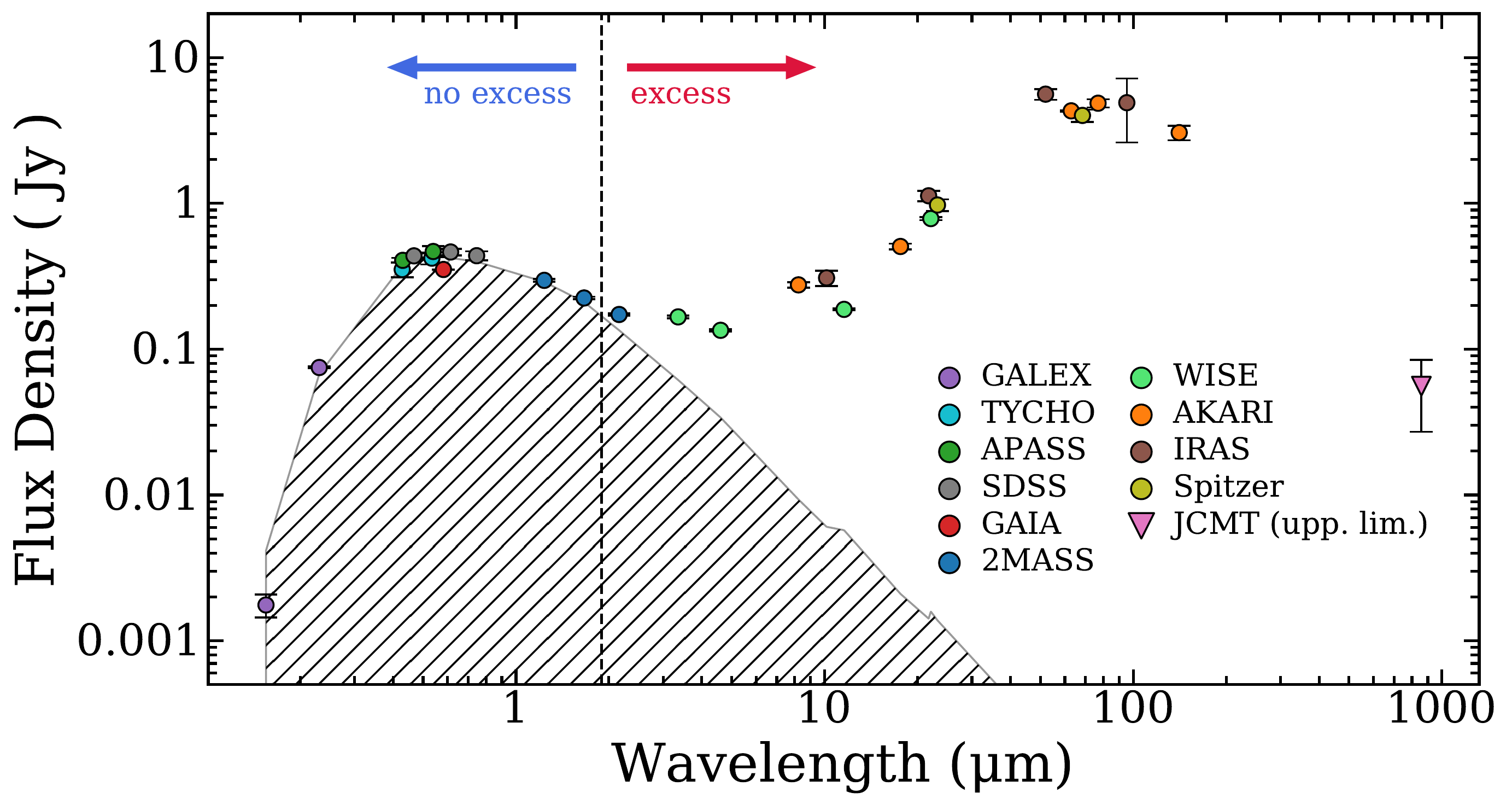}
\end{minipage}
\begin{minipage}{0.4\hsize}
    \centering
    \includegraphics[width=1\textwidth]{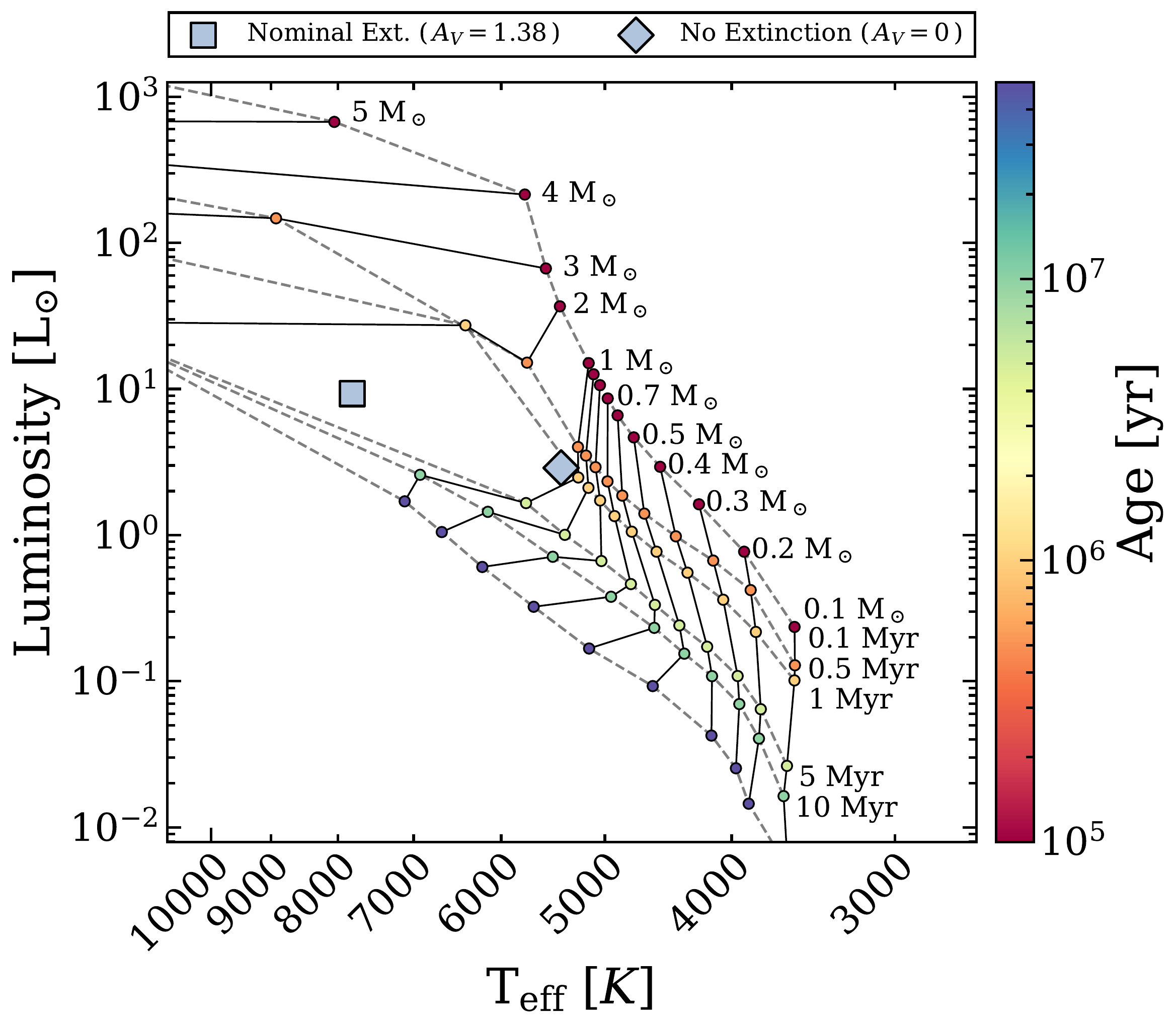}
\end{minipage}
\end{tabular}
    \caption{(Left) SED of J0337 overlaid with the best-fit stellar photosphere indicated by the gray hatched region. We used only the data points to the left of the vertical dashed line to characterize the stellar parameters. The points to the right of the vertical dashed line are regarded as the IR excess, the disk parameters for which are estimated in \cite{vanderMarel2016}. The inverted triangular point indicates an upper limit. The archival photometry is summarized in Appendix \ref{sec: J0337 SED}.
    (Right) Comparison of MIST evolutionary tracks (solid lines) and isochrones (dashed lines) with the best-fit parameters of J0337's SED, assuming either the nominal $A_{\rm V}$ value or no extinction. These parameters are consistent with a 1--5 Myr star, in agreement with the age estimate of the Perseus star forming region.}
    \label{fig: SED}
\end{figure*}

Due to lack of known spectral type, \cite{vanderMarel2016} adopted $A_{\rm v}=0.0$ but the Bayester19 dust map \citep{Green2019} suggests $A_{\rm v}\sim1.4$~mag around J0337, thus we corrected the photometry data from published survey data between UV and near-IR wavelengths. 
Figure \ref{fig: SED} shows J0337's SED ranging between UV and sub-mm wavelengths (see also Appendix \ref{sec: J0337 SED}), and we note that the previous surveys detected no flux at wavelengths longer than 200 $\micron$. 
For SED fitting, we permitted $A_{\rm v}$ values in the range $1.38 - 1.46$ (the 1-$\sigma$ range about the nominal value from the Bayestar19 dust map). Note that we do not account for circumstellar materials such as envelopes, if any, and in this sense the effective extinction may be underestimated.
We used available GALEX \citep{GALEX}, TYCHO \citep{TYCHO}, Gaia \citep{Gaia}, SDSS \citep{SDSS}, APASS \citep{APASS}, and 2MASS \citep{2MASS} $JH$-band data points for the stellar characterizations (the best-fit parameters: $T_{\rm eff}\sim7800$~K, $\log\ g\sim5.0$, $L_\star\sim9.3\ L_\odot$, and $M_\star\sim1.4\ M_\odot$ for $A_{\rm v} = 1.38$).
SED retrieval, dereddening, and fitting were carried out using the Virtual Observatory SED Analyzer \citep[VOSA;][]{Bayo2008}. 
For SED fitting, we utilized the BT-Settl-AGSS2009 synthetic stellar spectra \citep{Allard2012,Asplund2009}. The VOSA ``Chi-square Fit" tool was used to determine the best-fit parameter values provided above.
However, results from other VOSA SED fitting routines provide comparable solutions (e.g., the ``Model Bayes Analysis" indicates a 1-$\sigma$ confidence interval for $T_{\rm eff}$ in the range $7444 - 7800 K$). For comparison, we also performed fitting for $A_{\rm v} = 0$ \citep[no extinction;][]{vanderMarel2016}; in either case, the derived best-fit parameters are well located in the 1-5~Myr range of the MIST evolutionary models \citep[see the left panel of Fig \ref{fig: SED}, and for details see][]{Dotter2016}.
For further characterizations of the stellar parameters such as accretion high-dispersion spectroscopy is helpful \citep[][]{Manara2014,Alcala2017}.

We also extrapolated the disk parameters using the updated Gaia-based distance of 303~pc \citep{Gaia-DR2}.
As mentioned in \cite{vanderMarel2016} the estimated disk mass has large uncertainty because the SED fitting at wavelengths shorter than millimeter can trace only optically-thick part of the disk.
We use an upper limit of the JCMT/SCUBA-2 850~\micron\ observations, which is converted into $<20 M_{\rm Jup}$
with Equation (5) in \cite{vanderMarel2016}, for the discussion of the disk evolution mechanism. 
In Section \ref{sec: Forward Modeling} we also discuss the radiative transfer modeling to reproduce the outer arc feature.
However, we applied ADI technique to remove stellar halo and to detect the disk scattered light, which attenuates and distorts the disk feature.
Therefore in this study we focus on the detection of the spatially resolved large cavity. 
The detailed modeling will be performed with the follow-up polarimetric differential imaging \citep[PDI;][]{Kuhn2001} or RDI data as well as (sub-)millimeter observations of the J0337 disk.

\subsection{Forward Modeling} \label{sec: Forward Modeling}

To validate the cavity size we forward-modeled the disk feature to the NIRC2 data using RADMC-3D radiative transfer code \citep{RADMC3D} and {\tt pyklip} forward modeling modules. 
As mentioned in Section \ref{sec: SED Fitting}, however, it is difficult to estimate the detailed disk parameters with the current SED data points and we aim at reproducing the NIRC2 image and the MIR excess of the SED in the forward modeling processes. For simplicity we fix the inclination$=60^\circ$ and position angle$=-25^\circ$ in the modeling.
We found that the inner wall of the outer disk needs to have low surface density so that the forward scattering is brighter than the backward scattering with inclination$\sim60^\circ$, which can reproduce the NIRC2 results, on the other hand the outer disk needs to have a high enough dust mass (high surface density) to reproduce the MIR excess.
A power-law density profile ($\Sigma_{\rm dust} \propto r^{-0.5}$, where $\Sigma_{\rm dust}$ and $r$ correspond to the dust surface density and separation from the star respectively) can reproduce either of these two characteristics but cannot reproduce both simultaneously. 
Therefore, for reproducing both characteristics we utilized a Gaussian profile 
\begin{eqnarray}
\Sigma_{\rm dust}(r)=\Sigma_{\rm center}*\exp{(-(r-r_{\rm center})^2/2\sigma_{r}^2)},
\end{eqnarray}
where $\Sigma_{\rm center}=5\times10^{-3}$~g~cm$^{-2}$, $r_{\rm center}$= 200~au, and $\sigma_r$= 50~au; note that these parameters are degenerate). In this paper, we fix $\Sigma_{\rm center}$, $r_{\rm center}$, and $\sigma_r$ since near-IR does not in general trace dust surface density. The dust surface density distribution will be better constrained by spatially resolved sub-mm observations with e.g., ALMA. 

\begin{figure*}
\begin{tabular}{ccc}
\begin{minipage}{0.4\hsize}
    \centering
    \includegraphics[width=0.9\textwidth]{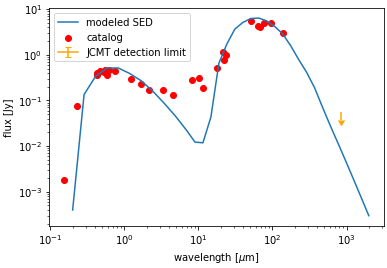}
\end{minipage}
\begin{minipage}{0.3\hsize}
    \centering
    \includegraphics[width=\textwidth]{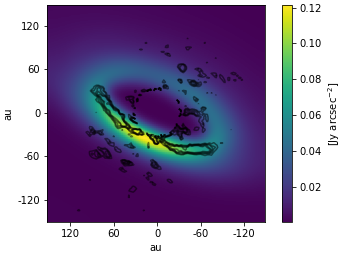}
\end{minipage}
\begin{minipage}{0.3\hsize}
    \centering
    \includegraphics[width=\textwidth]{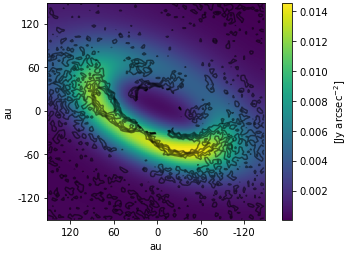}
\end{minipage}
\end{tabular}
    \caption{The adopted model assuming the cavity size of 80~au. (Left) The modeled SED in this study compared with the catalog values - we include only the stellar component with $T_{\rm eff}=7800$~K blackbody and the outer disk component (see text). The discrepancy at $\sim10\micron$ suggests an unseen inner disk. (Middle) The modeled disk image at $K_{\rm p}$ band overlaid with the NIRC2 contours. (Right) Same as the middle panel for $L_{\rm p}$ band. Note that the modeled disk images are smoothed by Gaussian with the measured FWHM of each NIRC2 observation.}
    \label{fig: best model}
\end{figure*}

\begin{figure*}
\begin{tabular}{cc}
\begin{minipage}{0.5\hsize}
    \centering
    \includegraphics[width=0.8\textwidth]{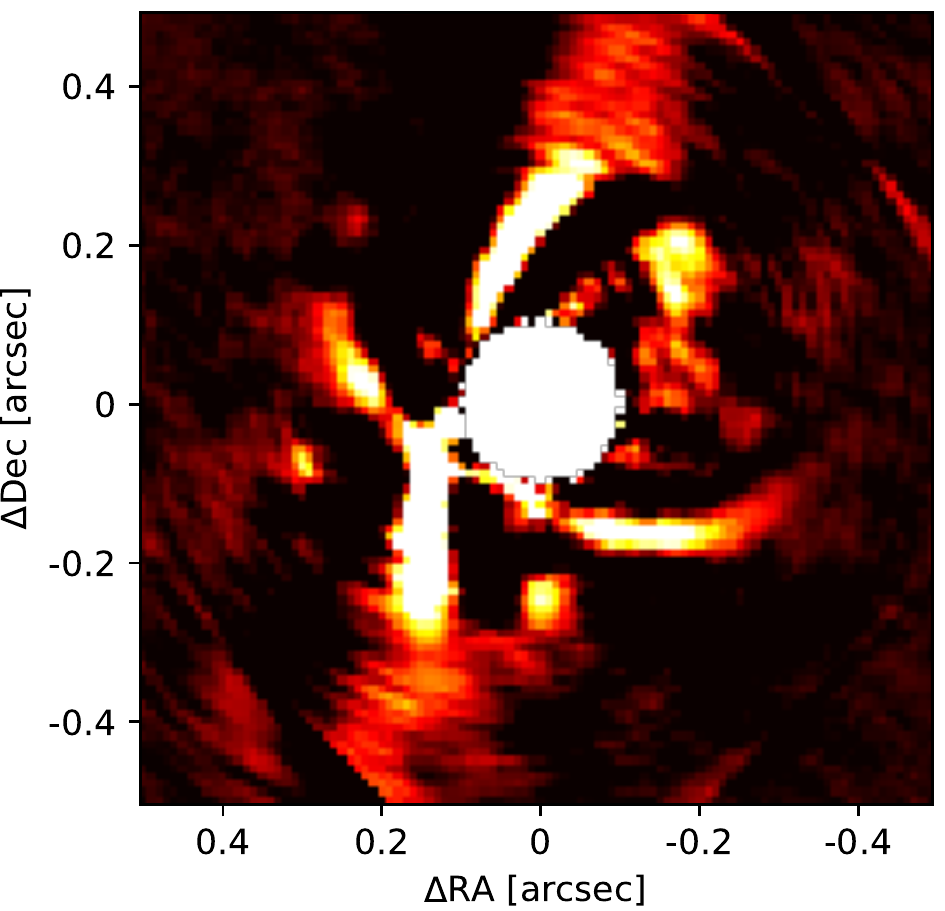}
\end{minipage}
\begin{minipage}{0.5\hsize}
    \centering
    \includegraphics[width=0.8\textwidth]{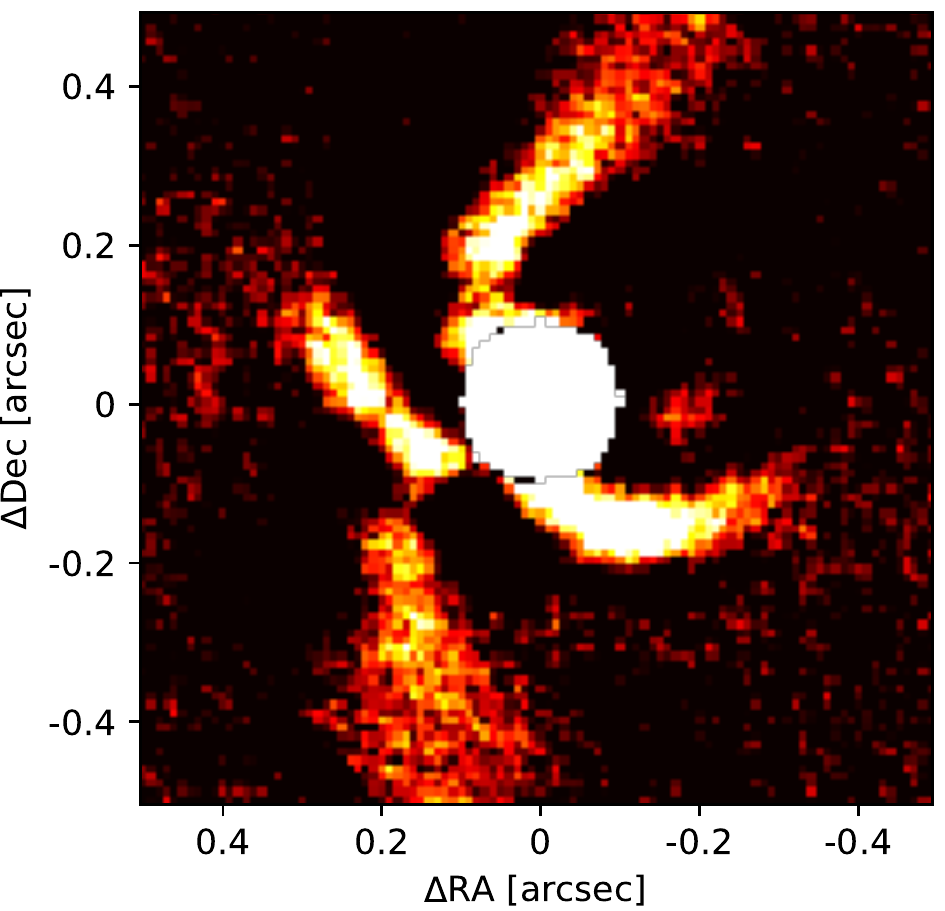}
\end{minipage}
\end{tabular}
    \caption{Post-processed images at $K_{\rm p}$ (left) and $L_{\rm p}$ band (right) after injecting the 80au-cavity model (Figure \ref{fig: best model}) at position angle of -115$^\circ$.}
    \label{fig: best forwarded model}
\end{figure*}

We also adopted a dust settling factor $f_{\rm set}=2$ to match the observational results and a scale height is calculated as $\frac{\sqrt{kT_{\rm dust}/(2.34m_{\rm H})}}{f_{\rm set}\sqrt{GM_\star/r^3}}$, where $k, T_{\rm dust}, m_{\rm H}, G$ are the Boltzmann constant, dust temperature, mass of hydrogen, and gravitational constant, respectively. We used thermal Monte Carlo computations of RADMC-3D (mctherm) to calculate the dust temperature derived from the adopted dust density profile and the stellar parameters.
We then utilized the Mie scattering codes \citep{Bohren1983} with the Gaussian dust-size distribution (centered around 10~$\micron$, 5\% width) and the optical constant of amorphous silicate\footnote{\url{http://www.astro.uni-jena.de/Laboratory/OCDB/data/silicate/amorph/pyrmg70.lnk}} to calculate dust opacity. 
These assumptions provide a good match to the catalog values around MIR and we tested the cavity size by defining a cutoff of the modeled disk ($\Sigma_{\rm dust}(r)=0$ at $r<r_{\rm cutoff}$ in Equation 1) as the cavity radius and changing it from 40~au to 100~au by a 20-au step. 
We injected the modeled disk to the NIRC2 data at a position angle of -115$^\circ$ (rotated by 90$^\circ$ clockwise) and then rereduced them. 
The 80-au cavity model seems to best match the NIRC2 results while the SEDs are not largely affected within the parameter range of the cavity radius we explored (see Figures \ref{fig: best model} and \ref{fig: best forwarded model} and Appendix \ref{sec: Radmc3D Modeling}). Note that because of the imperfect AO correction and the presence of the actual disk feature, the injected disk feature is affected to some extent in the post-processed image. The $L_{\rm p}$-band forward-modeled images are less clear than the $K_{\rm p}$-band images as the $L_{\rm p}$-band observations have the larger PSF (see also Section \ref{sec: Observations}). Particularly the southern part of the injected disk is attenuated by self-subtraction of the actual disk features.
In Table \ref{tab: adopted parameters} we summarize the stellar and disk parameters we adopted in this study. 
Since the system is yet relatively unexplored and we only have images in near-IR wavelengths, it is hard determine all the model parameters. 
We therefore address the uncertainties of the model parameters briefly by changing stellar luminosity, whose uncertainty is expected to be a factor of a few due to the uncertainty of extinction towards the star. The change of the stellar luminosity affects the disk temperature. As shown in Appendix \ref{sec: Radmc3D Modeling - test dust}, we see that the modeled SED is affected by the stellar luminosity while the geometry of the modeled near-IR image is not. Therefore, we consider that the cavity radius of $\sim80$~au is relatively robust despite the uncertainty of stellar luminosity.

We also note that our modeled SED cannot reproduce the $\sim10\micron$ excess and the discrepancy indicates presence of an unseen inner disk. 
As the NIRC2 data did not confirm any inner disk features the inner disk may be located within or close to the inner working angle of the NIRC2 observations ($\sim0\farcs1=30\ {\rm au}$), 
which is indeed well outside the typical radii of inner disks \citep{Francis2020}.
Future interferometric observations or follow-up PDI/RDI observations will be able to discuss the inner disk. Particularly the follow-up PDI/RDI may detect a shadowing effect by the inner disk \citep[e.g.][]{Bohn2021} and this should be taken into account for a better model of the J0337 disk.

\begin{table}
    \caption{Summary of the adopted parameters}
    \label{tab: adopted parameters}
    \centering
    \begin{tabular}{cc} \hline\hline
     parameter  & value  \\ \hline
     \multicolumn{2}{c}{Stellar parameters} \\ 
     $T_{\rm eff}$ & 7800~K \\
     $\log g$ & 5.0 \\
     $L_\star$ & 9.3 $L_\odot$  \\
     $M_\star$ & 1.4 $M_\odot$   \\
     age & 1-5~Myr \\
     $A_{\rm V}$ & 1.38  \\ \hline
     \multicolumn{2}{c}{Disk parameters} \\ 
     inclination & 60$^\circ$  \\
     position angle & -25$^\circ$ \\
     dust distribution & Gaussian profile (Eqn. 1) \\
     cavity radius & 80~au \\
    \end{tabular}
\end{table}

\subsection{Gap Opening Mechanism}

\cite{Alexander2006} simulated the disk evolution with photoevaporation and predicted that the disk with a photoevaporation-induced cavity has the order of $M_{\rm disk}\sim10^{-4}\ M_\star$. 
With the updated stellar parameters and the upper limit of the disk mass (see Section \ref{sec: SED Fitting}), the disk mass is $<1.4\times10^{-2}\ M_\star$. If the actual disk mass is smaller than the upper limit by an order of magnitude photoevaporation can be responsible for the cavity, but it is unlikely that photoevaporation is the only mechanism that opens such a large cavity \citep[e.g.][]{Espaillat2010}. Accretion rate is another index to discuss the mass loss ratio by photoevaporation \citep[e.g.][]{Owen2012, Ercolano2017}, but J0337's accretion has not been studied.
Deep radio-wavelengths explorations and high-dispersion spectroscopy for this system will help to determine the disk mass and to further investigate the photoevaporation scenario.

\begin{figure*}
\begin{tabular}{cc}
\begin{minipage}{0.5\hsize}
    \centering
    \includegraphics[width=0.8\textwidth]{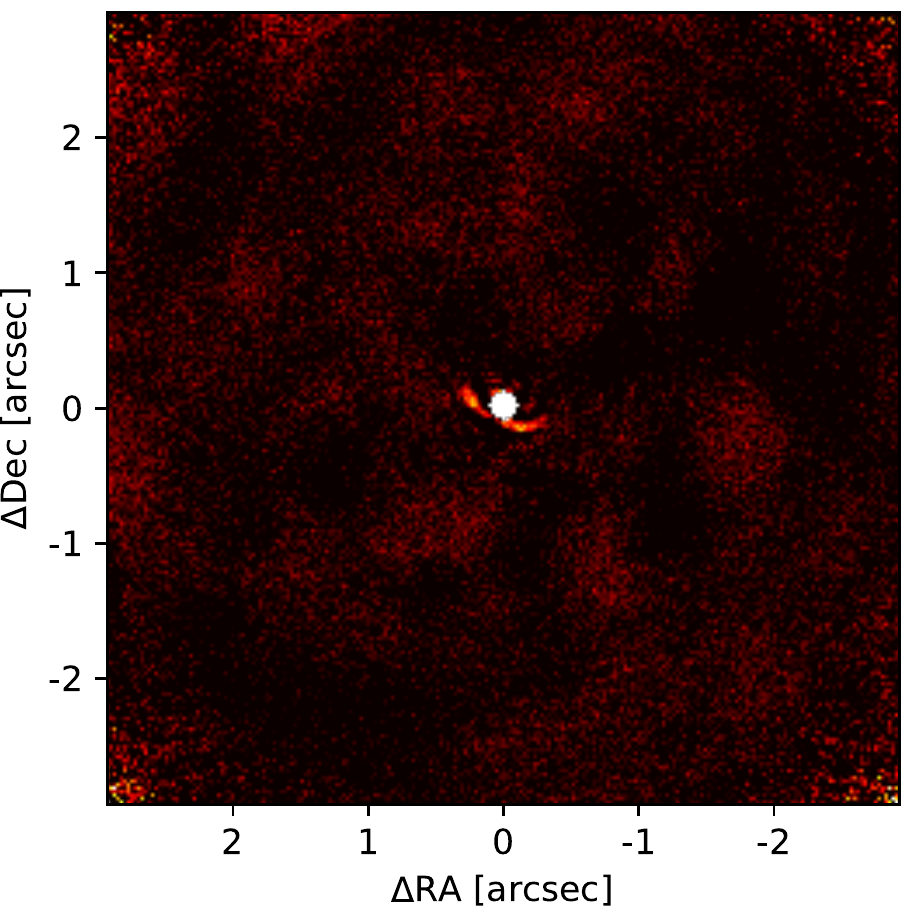}
    \caption{Same NIRC2 $L_{\rm p}$-band result (see Figure \ref{fig: NIRC2 result}) with the larger FoV and KL=10 to search for outer companion candidates.}
    \label{fig: NIRC2 large FoV}
\end{minipage}
\begin{minipage}{0.5\hsize}
    \centering
    \includegraphics[width=0.95\textwidth]{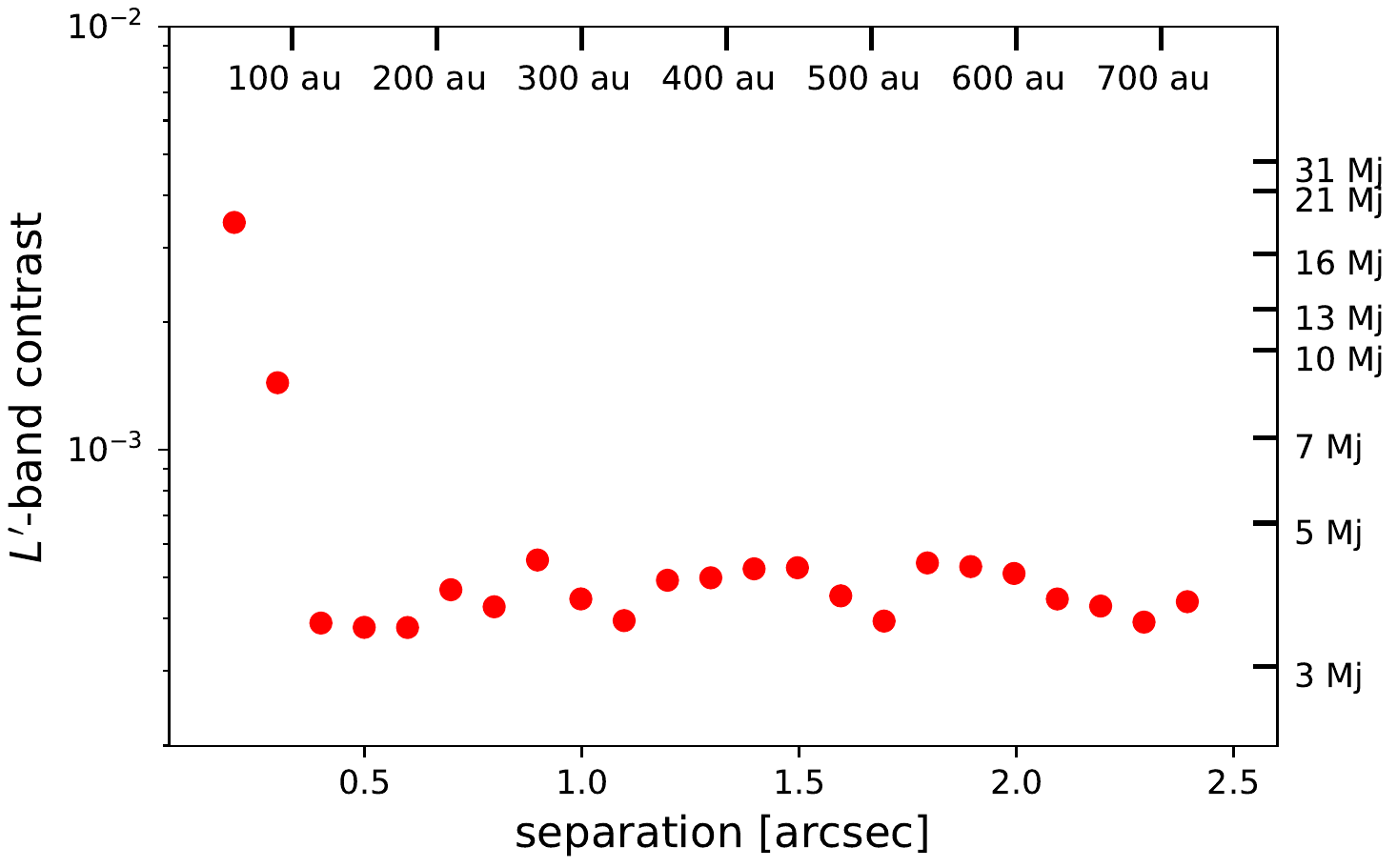}
    \caption{5$\sigma$ detection limit of the $L_{\rm p}$-band observations made from the KL=10 result (see Figure \ref{fig: NIRC2 large FoV}).}
    \label{fig: contrast}
\end{minipage}
\end{tabular}
\end{figure*}

Planet formation is also one of the most plausible scenarios given that the cavity size is similar to the PDS~70 system \citep{Keppler2018}.
However, our observations did not detect any convincing companion candidates at separations between $0\farcs1$ and $2\farcs5$. Figure \ref{fig: NIRC2 large FoV} shows the $L_{\rm p}$-band ADI result with KL=10 to search for companions. Figure \ref{fig: contrast} shows 5$\sigma$ detection limit of Figure \ref{fig: NIRC2 large FoV} with the expected contrast of a substellar-mass object assuming 1 Myr and COND03 model \citep{Baraffe2003}. We divided the FoV into a number of annular regions and defined the noise as the standard deviation within each annular region, where we masked the disk feature and used 5$\sigma$ clipping to mitigate the effect of the disk feature on the detection limit. We also corrected the self-subtraction effect by injecting fake sources at different position angle and separations and calculating the flux-attenuation ratio.
Out detection limits could constrain the presence of a massive brown-dwarf companion in the cavity region ($\sim20\ M_{\rm Jup}$ at 60~au and $\sim9-10\ M_{\rm Jup}$ at 90~au) and that of a $\sim3\ M_{\rm Jup}$ protoplanet outside the cavity ($>120$~au).
Compared with other high-contrast imaging and simulation studies that targeted large-cavity transitional disks such as PDS~70 or RX~J1604 \citep{Haffert2019,Muley2019,Canovas2017} there could be multiple Jovian planets less than $10\ M_{\rm Jup}$, a single eccentric Jovian planet, or a brown-dwarf companion close to the central star that we cannot resolve in our observations.

\section{Summary} \label{sec: Summary}
 
YSOs in Perseus are typically more distant than other nearby star forming regions such as Taurus, Lupus, and Ophiuchus. Furthermore Perseus is located at northern in the celestial field and Perseus has been less prioritized by ALMA or adaptive optics observations.
In this study we presented Keck/NIRC2 $K_{\rm p}L_{\rm p}$ high-contrast imaging observations of the J0337 protoplanetary disk, one of the transitional disk candidates with a large cavity suggested in \cite{vanderMarel2016}. 
After ADI reduction with {\tt pyklip} we detected forward scattering from an inner wall of the outer disk and affirmed the large cavity. As the ADI reduction distorts the geometry of the disk, we investigated the cavity size with forward-modeled disks that are made from RADMC-3D radiative transfer codes as well as aiming to reproduce its SED. Our data and forward modeling suggest $\sim80$~au for the cavity radius, which is consistent with the prediction of \cite{vanderMarel2016}. However there is discrepancy at $\sim10\micron$ in J0337's SED and the modeled SED suggesting the presence of an unseen inner disk.
We also searched for companions around J0337 but did not detect any convincing companions within $2\farcs5$ and the $L_{\rm p}$-band detection limit could set a constraint on the mass of potential companions to $\sim20 M_{\rm Jup}$ at 60~au, $\sim9-10 M_{\rm Jup}$ at 90~au, and $\sim3 M_{\rm Jup}$ at $>120$~au assuming 1~Myr and COND03~model. Compared with other protoplanetary disk systems with large cavities such as PDS~70 and RX~J1604, a massive brown-dwarf companion at an inner separation or multiple Jovian planets fainter than our detection limit within the cavity may exist inside the cavity.

The detailed modeling, including the inner disk component, will help better understand the J0337 system while follow-up observations such as spectroscopic observation of J0337 for the stellar characterization, PDI/RDI to investigate the scattered light without self-subtraction, obtaining radio flux, and ALMA high-angular resolution observation are essential.
Finally our discovery is the second spatially resolved cavity in the Perseus protoplanetary disks and there remain more transitional disks suggested in SED-based studies such as \cite{vanderMarel2016}, which promotes further high angular resolution observations at this region.

\newpage
\acknowledgments
The authors would like to thank the anonymous referees for their constructive comments and suggestions to improve the quality of the paper.
We thank Akimasa Kataoka for helpful comments to develop the RADMC-3D modeling.
T.U. is supported by Grant-in-Aid for Japan Society for the Promotion of Science (JSPS) Fellows and JSPS KAKENHI Grant No. JP21J01220.
T. M. is supported by JSPS KAKANHI Grant Nos. 17H01103, 18H05441 and 19K03932.
The data presented herein were obtained at the W. M. Keck Observatory, which is operated as a scientific partnership among the California Institute of Technology, the University of California and the National Aeronautics and Space Administration. The Observatory was made possible by the generous financial support of the W. M. Keck Foundation.
The authors wish to recognize and acknowledge the very significant cultural role and reverence that the summit of Maunakea has always had within the indigenous Hawaiian community.  We are most fortunate to have the opportunity to conduct observations from this mountain.
This research has made use of the Keck Observatory Archive (KOA), which is operated by the W. M. Keck Observatory and the NASA Exoplanet Science Institute (NExScI), under contract with the National Aeronautics and Space Administration.
This publication makes use of VOSA, developed under the Spanish Virtual Observatory project supported by the Spanish MINECO through grant AyA2017-84089.
VOSA has been partially updated by using funding from the European Union's Horizon 2020 Research and Innovation Programme, under Grant Agreement nº 776403 (EXOPLANETS-A)

\appendix
\section{J0337 SED} \label{sec: J0337 SED}
Table \ref{tab: SED J0337} summarizes archival photometry of J0337 (see also Figure \ref{fig: SED}). We used these catalog values no longer than 1.7$\micron$ for the SED fitting to estimate the stellar parameters and referred to the MIR photometry for the forward modeling.

\begin{table*}
\centering
\caption{Archival photometry of J0337}
\begin{tabular}{cccc} \hline\hline
wavelength [\micron] & flux [erg~s$^{-1}$~cm$^{-2}$~\micron$^{-1}$] & flux error [erg~s$^{-1}$~cm$^{-2}$~\micron$^{-1}$]  & remarks \\ \hline
0.155 & 2.2e-10 & 3.97e-11 & GALEX \citep{GALEX} \\
0.23 & 4.23e-09 & 6.46e-11 & GALEX \citep{GALEX} \\
0.428 & 5.75e-09 & 6.63e-10 & TYCHO \citep{TYCHO} \\
0.43 & 6.60e-09 & 2.61e-10 & APASS \citep{APASS} \\
0.467 & 6.00e-09 & 2.65e-10 & SDSS \citep{SDSS} \\
0.534 & 4.43e-09 & 4.24e-10 & TYCHO \citep{TYCHO} \\
0.539 & 4.82e-09 & 4.22e-10 & APASS \citep{APASS} \\
0.582 & 3.11e-09 & 8.17e-12 & Gaia \citep{Gaia} \\
0.614 & 3.69e-09 & 1.87e-10 & SDSS \citep{SDSS} \\
0.746 & 2.36e-09 & 1.67e-10 & SDSS \citep{SDSS} \\
1.235 & 5.83e-10 & 1.23e-11 & 2MASS \citep{2MASS} \\
1.662 & 2.44e-10 & 4.71e-12 & 2MASS \citep{2MASS} \\
2.159 & 1.11e-10 & 1.74e-12 & 2MASS \citep{2MASS} \\
3.353 & 4.44e-11 & 9.41e-13 & WISE \citep{WISE} \\
4.603 & 1.91e-11 & 3.51e-13 & WISE \citep{WISE} \\
8.228 & 1.22e-11 & 5.39e-13 & AKARI \citep{AKARI} \\
10.146 & 8.98e-12 & 1.08e-12 & IRAS \citep{IRAS} \\
11.561 & 4.21e-12 & 6.2e-14 & WISE \citep{WISE} \\
17.609 & 4.89e-12 & 2.26e-13 & AKARI \citep{AKARI} \\
21.727 & 7.16e-12 & 5.85e-13 & IRAS \citep{IRAS} \\
22.088 & 4.83e-12 & 1.11e-13 & WISE \citep{WISE} \\
23.21 & 5.42e-12 & 5.04e-13 & Spitzer \citep{Spitzer} \\
51.989 & 6.22e-12 & 5.24e-13 & IRAS \citep{IRAS} \\
62.951 & 3.26e-12 & 3.6e-14 & AKARI \citep{AKARI} \\
68.445 & 2.57e-12 & 2.48e-13 & Spitzer \citep{Spitzer} \\ 
76.904 & 2.46e-12 & 1.62e-13 & AKARI \citep{AKARI} \\
95.297 & 1.62e-12 & 7.56e-13 & IRAS \citep{IRAS} \\
140.856 & 4.62e-13 & 5.26e-14 & AKARI \citep{AKARI} \\
857.914 & 2.27e-16 & 1.17e-16 & JCMT upper limit \citep{vanderMarel2016}\\
\end{tabular}
    \label{tab: SED J0337}
\end{table*}

\section{Forward Modeling with Different Cavity Sizes} \label{sec: Radmc3D Modeling}

\begin{figure*}
\begin{tabular}{cccc}
\begin{minipage}{0.25\hsize}
    \centering
    \includegraphics[width=\textwidth]{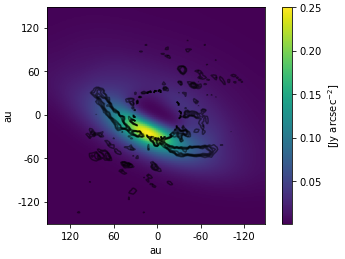}
\end{minipage}
\begin{minipage}{0.25\hsize}
    \centering
    \includegraphics[width=0.8\textwidth]{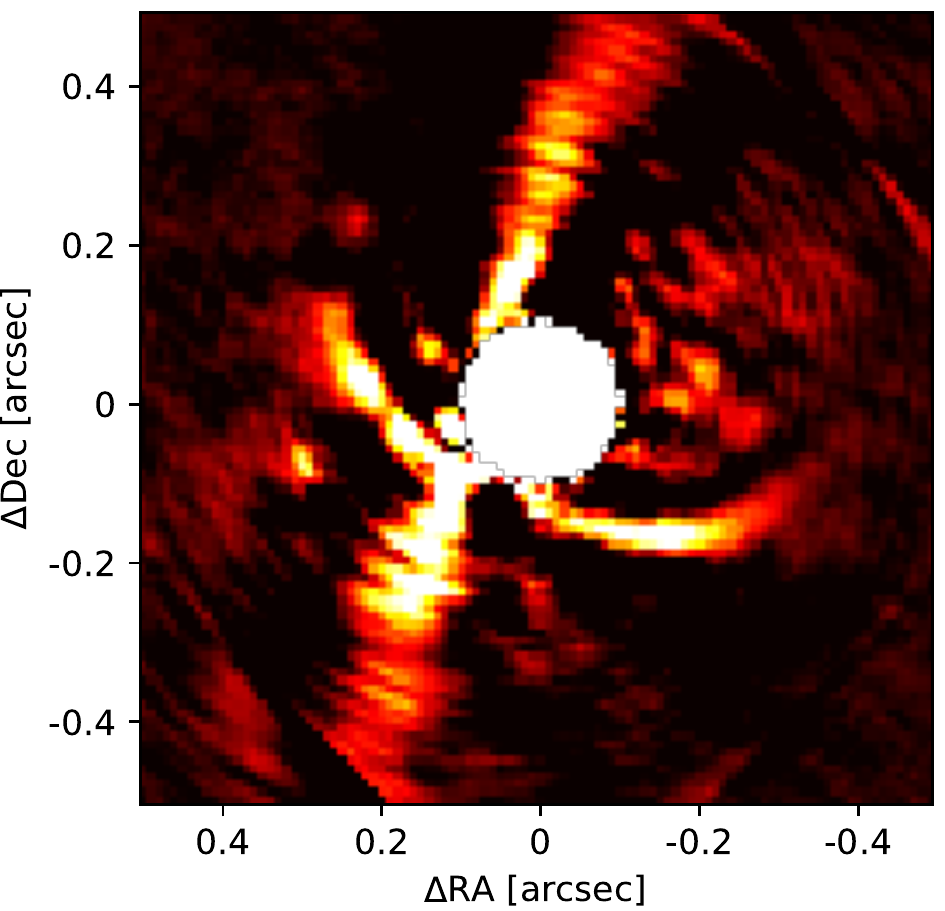}
\end{minipage}
\begin{minipage}{0.25\hsize}
    \centering
    \includegraphics[width=\textwidth]{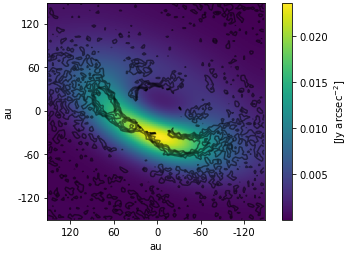}
\end{minipage}
\begin{minipage}{0.25\hsize}
    \centering
    \includegraphics[width=0.8\textwidth]{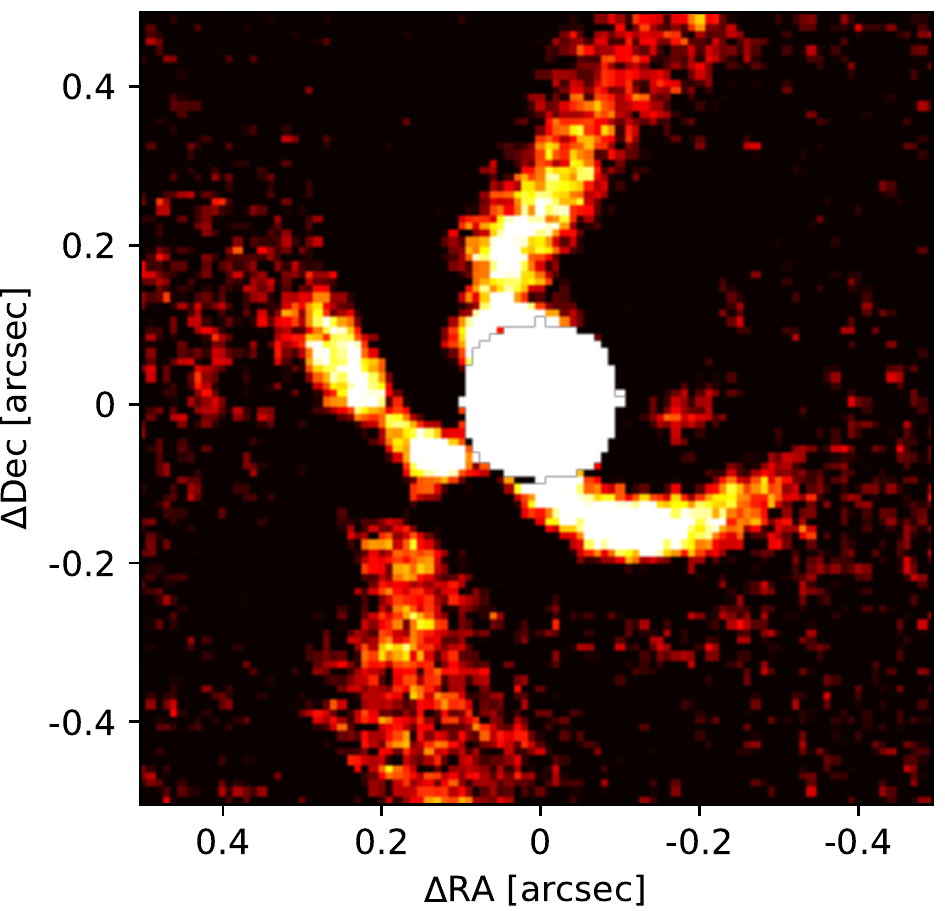}
\end{minipage}
\end{tabular}
    \caption{The modeled disk images assuming the cavity size of 40~au. From left to right: modeled disk at $K_{\rm p}$ band overlaid with the NIRC2 result contours (black), post-processed NIRC2 image at $K_{\rm p}$ band after injecting the modeled disk (forward-modeled result), modeled disk at $L_{\rm p}$ band, and forward-modeled result at $L_{\rm p}$ band.}
    \label{fig: model 40au}
\end{figure*}

\begin{figure*}
\begin{tabular}{cccc}
\begin{minipage}{0.25\hsize}
    \centering
    \includegraphics[width=\textwidth]{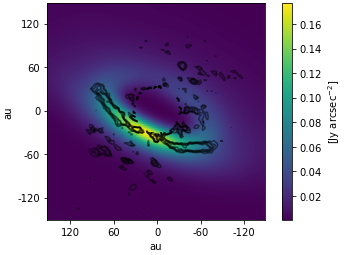}
\end{minipage}
\begin{minipage}{0.25\hsize}
    \centering
    \includegraphics[width=0.8\textwidth]{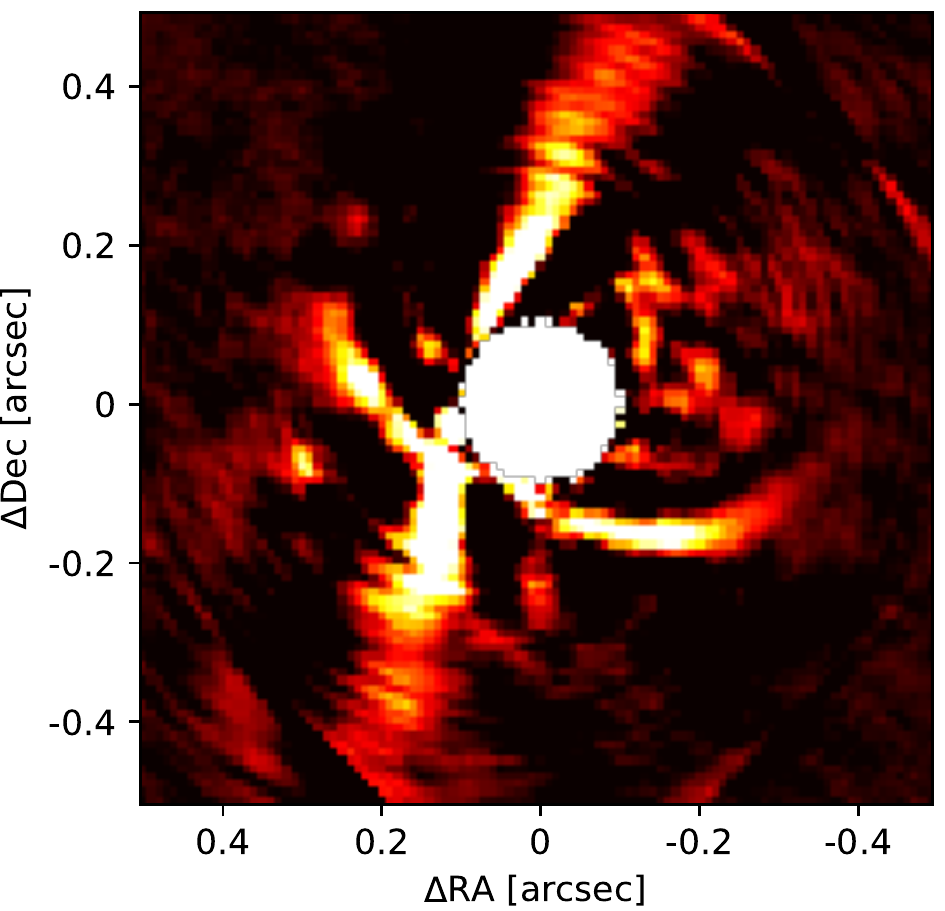}
\end{minipage}
\begin{minipage}{0.25\hsize}
    \centering
    \includegraphics[width=\textwidth]{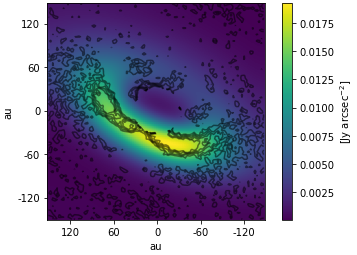}
\end{minipage}
\begin{minipage}{0.25\hsize}
    \centering
    \includegraphics[width=0.8\textwidth]{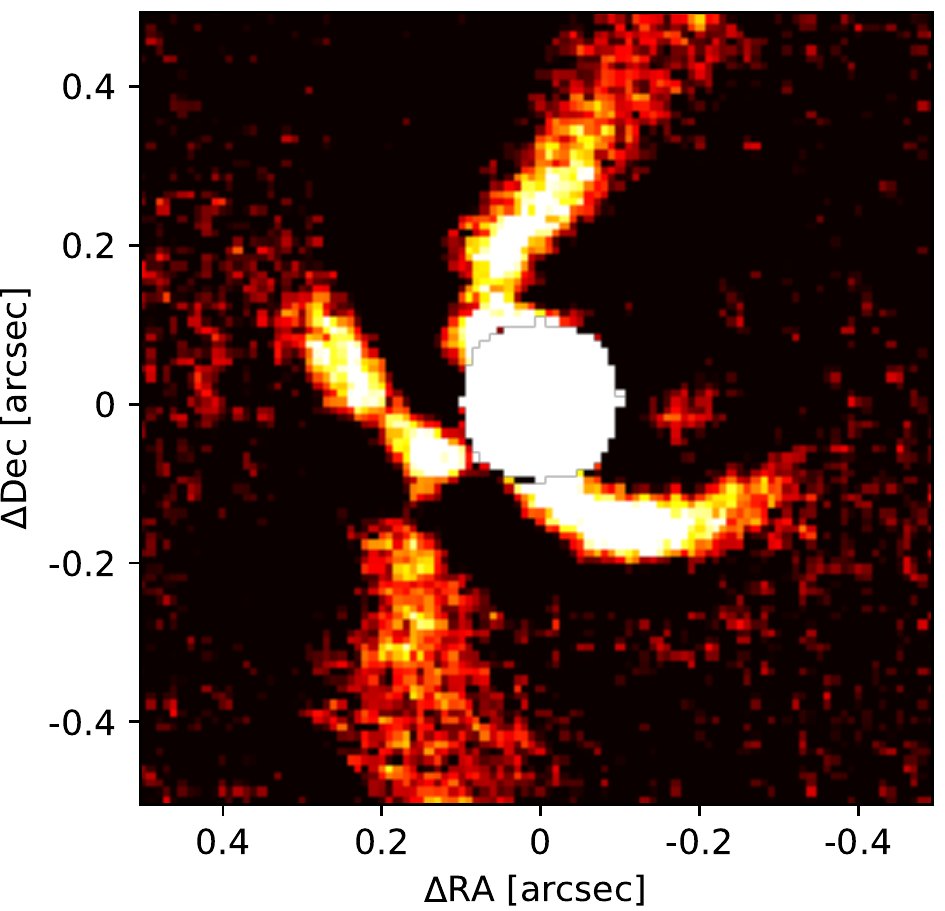}
\end{minipage}
\end{tabular}
    \caption{Same as Figure \ref{fig: model 40au} for the case of 60au-cavity.}
    \label{fig: model 60au}
\end{figure*}

\begin{figure*}
\begin{tabular}{cccc}
\begin{minipage}{0.25\hsize}
    \centering
    \includegraphics[width=\textwidth]{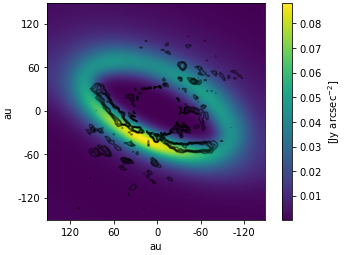}
\end{minipage}
\begin{minipage}{0.25\hsize}
    \centering
    \includegraphics[width=0.8\textwidth]{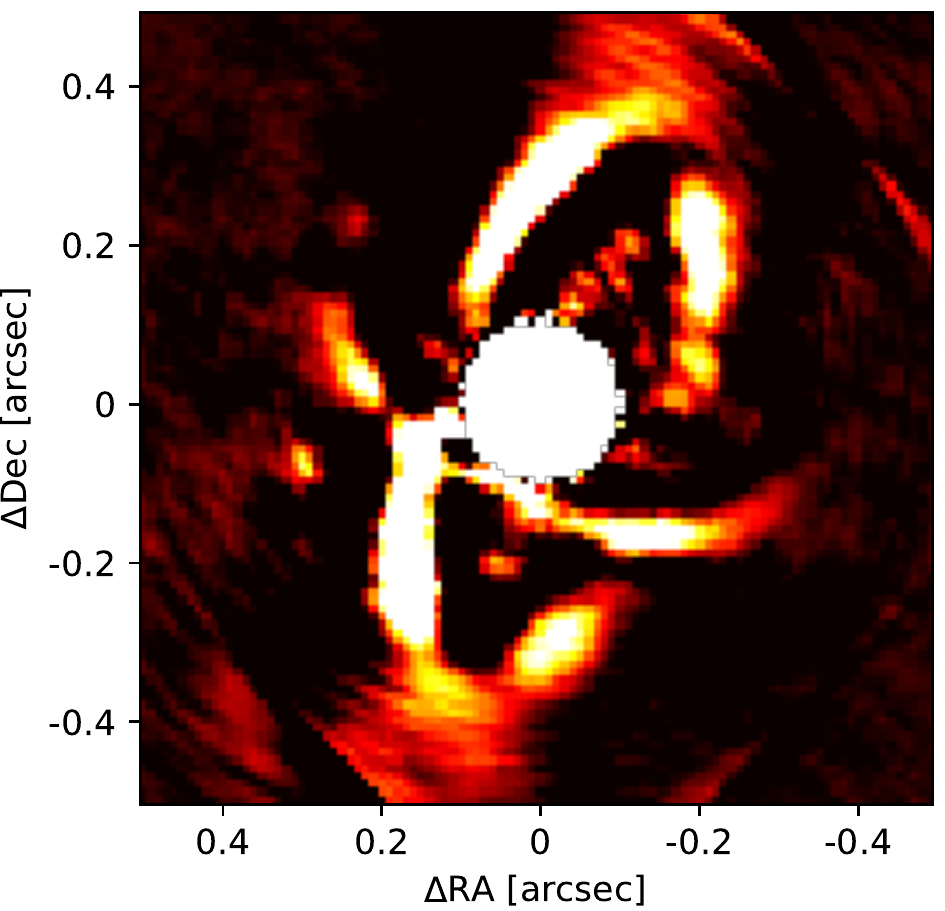}
\end{minipage}
\begin{minipage}{0.25\hsize}
    \centering
    \includegraphics[width=\textwidth]{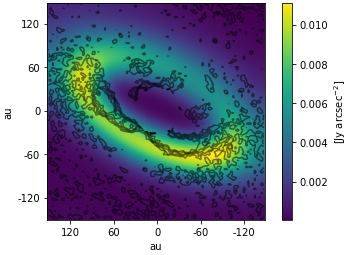}
\end{minipage}
\begin{minipage}{0.25\hsize}
    \centering
    \includegraphics[width=0.8\textwidth]{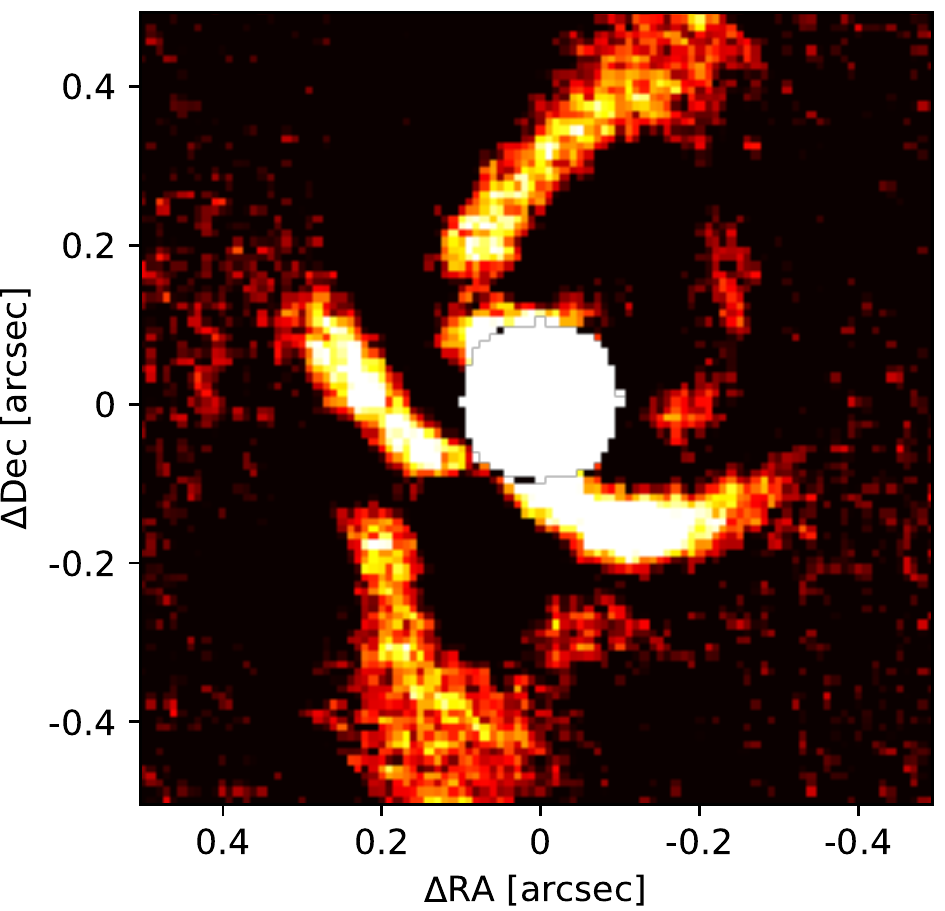}
\end{minipage}
\end{tabular}
    \caption{Same as Figures \ref{fig: model 40au}, \ref{fig: model 60au} for the case of 100au-cavity.}
    \label{fig: model 100au}
\end{figure*}

Figures \ref{fig: model 40au}, \ref{fig: model 60au}, and \ref{fig: model 100au} compare the model images and the forward-modeled results assuming 40, 60, and 100~au for the cavity size (defined as a cutoff in the model) in radius respectively. The Gaussian radial profile for the outer disk is fixed in the modeling. Regarding the scattering profile we multiplied the $K_{\rm p}$ and $L_{\rm p}$ modeled disks by 0.8 and 2 to correct the difference between the simulated blackbody with $T_{\rm eff}=7800$~K and the catalog values.
Considering the geometry of the disk scattered light, the 80au-cavity model best matches the NIRC2 results.
In the $K_{\rm p}$-band forward-modeled images, the 40au- and 60au-cavity models are reproduced as an almost straight line because the ADI reduction distorted the modeled disk, while the NIRC2 data show the arc feature as presented in Section \ref{sec: Results}. The 100au-cavity model shows the larger ring feature than the NIRC2 data and suggests the bright backward-scattering feature that is not seen in fact.

\section{RADMC-3D Modeling with Different Dust Temperature} \label{sec: Radmc3D Modeling - test dust}

\begin{figure}
    \centering
    \includegraphics[width=0.5\textwidth]{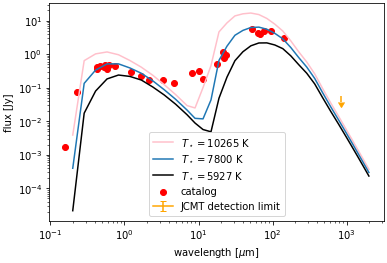}
    \caption{Comparison of the modeled SEDs assuming different temperature of the central star and the 80-au cavity. The SED with $T_\star=7800$~K is the same as Figure \ref{fig: best model}.}
    \label{fig: test temp sed}
\end{figure}

\begin{figure*}
\begin{tabular}{cccc}
\begin{minipage}{0.25\hsize}
    \centering
    \includegraphics[width=\textwidth]{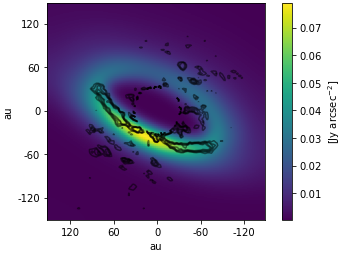}
\end{minipage}
\begin{minipage}{0.25\hsize}
    \centering
    \includegraphics[width=\textwidth]{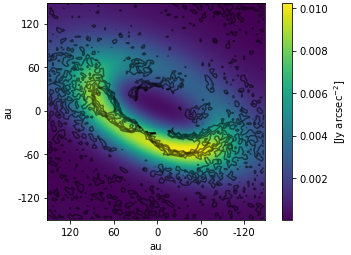}
\end{minipage}
\begin{minipage}{0.25\hsize}
    \centering
    \includegraphics[width=\textwidth]{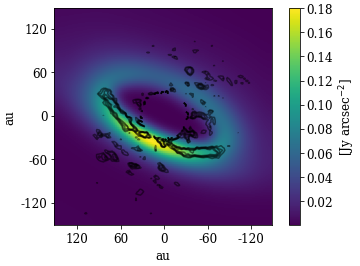}
\end{minipage}
\begin{minipage}{0.25\hsize}
    \centering
    \includegraphics[width=\textwidth]{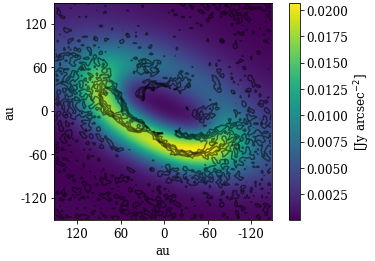}
\end{minipage}
\end{tabular}
    \caption{Comparison of the modeled images assuming different temperature of the central star and the 80-au cavity. From left to right: modeled disk at $K_{\rm p}$ band with $T_\star=5926$~K, modeled disk at $L_{\rm p}$ band with $T_\star=5926$, and modeled disk at $K_{\rm p}$ and $L_{\rm p}$ bands with $T_\star=10265$~K.}
    \label{fig: test temp image}
\end{figure*}

The uncertainty on luminosity from the VOSA fitting is 0.18 but as mentioned in Section \ref{sec: SED Fitting} the uncertainty on the extinction estimation could affect characterizing the stellar parameters. 
To test the effect of the dust temperature on the RADMC-3D modeling results, we varied a larger range of the stellar luminosity from 3.1~$L_\odot$ ($L_\star/3$) to 27.9~$L_\odot$ ($3L_\star$) by changing the stellar temperature in the RADMC-3D settings from 5927  ($7800/3^{0.25}$)~K to 10265 ($7800\times3^{0.25}$)~K while we fixed the cavity size to 80~au. The dust temperature was accordingly calculated by the 'mctherm' command in RADMC-3D.
Figure \ref{fig: test temp sed} shows the modeled SEDs and Figure \ref{fig: test temp image} compares the modeled images with the above assumptions.
The SED and surface brightness of the disk are variable with the stellar luminosity while the geometry of the disk feature does not vary within this luminosity range.

\bibliography{library}                                    
\end{document}